\documentclass[aps,prl,twocolumn,letterpaper,showpacs,superscriptaddress,floats]{revtex4}

\usepackage{graphics}
\usepackage{float}

\begin{document}
\setlength{\parskip}{0mm}
\preprint{}

\title{Multiple Time Scales in Diffraction Measurements of Diffusive Surface Relaxation}
\author{Aaron Fleet}
\affiliation{School of Applied and Engineering Physics, Cornell University, Ithaca, NY 14853}
\affiliation{Cornell Center for Materials Research, Cornell University, Ithaca, NY 14853}

\author{Darren Dale}
\affiliation{Department of Materials Science and Engineering, Cornell University, Ithaca, NY 14853}
\affiliation{Cornell Center for Materials Research, Cornell University, Ithaca, NY 14853}
\author{A.R. Woll}
\affiliation{Cornell High Energy Synchrotron Source, Cornell University, Ithaca, NY 14853}
\author{Y. Suzuki}
\affiliation{Department of Materials Science and Engineering, UC Berkeley, Berkeley, CA 94720}
\author{J. D. Brock}
\affiliation{School of Applied and Engineering Physics, Cornell University, Ithaca, NY 14853}
\affiliation{Cornell Center for Materials Research, Cornell University, Ithaca, NY 14853}

\date{\today}

\begin{abstract}
We grew SrTiO$_3$ on SrTiO$_3$ (001) by pulsed laser deposition, using x-ray scattering to monitor the growth in real time.  The time-resolved small angle scattering exhibits a well-defined length scale associated with the spacing between unit cell high surface features.  This length scale imposes a discrete spectrum of Fourier components and rate constants upon the diffusion equation solution, evident in multiple exponential relaxation of the ``anti-Bragg" diffracted intensity.  An Arrhenius analysis of measured rate constants confirms that they originate from a single activation energy.
\end{abstract}


\maketitle

Despite decades of study, a complete description of the fundamental mechanisms that control a broad class of epitaxial growth techniques persists as a challenging problem in non-equilibrium physics.  For example, in the thin film-growth technique of Pulsed Laser Deposition (PLD), an extremely dense plume of laser-ablated material deposits a high concentration of ionic and neutral species onto an atomically smooth surface~\cite{willmott_00}.  Numerous excellent diffraction-based studies~\cite{lippmaa_apl,karl,blank,wang} of the ensuing kinetics have revealed multiple time scales in the post-pulse intensity transient, and have suggested that two or more distinct physical mechanisms govern the surface's relaxation toward equilibrium.  In this Letter, we combine elementary diffusion theory and kinematic scattering theory with experiment to show that a \emph{single} energy process can produce a non-exponential diffraction response during PLD.  A better understanding of surface kinetics may explain why PLD can produce atomically smooth films, rivalling other growth techniques (e.g. molecular-beam epitaxy, chemical vapor deposition).  The sensitivity of functional properties to smoothness necessitates this understanding, since PLD is a candidate to grow device-quality films of complex-oxide materials \cite{ohtomo_04,viret,lowndes_96}.

Film-growth models typically assume that particles diffuse on the substrate until they either evaporate, attach to an existing step edge, or run into other particles and nucleate islands~\cite{BCF,petrich}.  Depending on the substrate temperature and specific energy barriers, films grow in one of several well-known modes: 3D, layer-by-layer, or step-flow~\cite{metev,neave,song}. These broad categories encompass myriad film morphologies, and the specific morphology determines the form of the intensity in diffraction studies of film growth.

\begin{figure}
\resizebox{85mm}{!}{\includegraphics{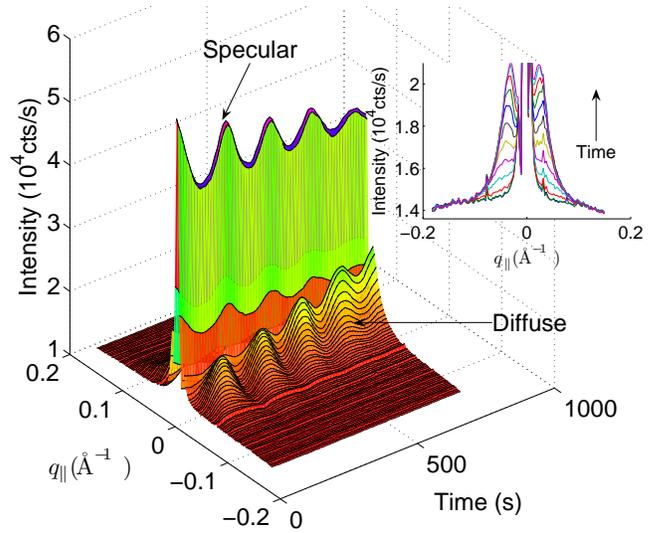}}

\caption{
X-ray intensity measured around $(00\frac{1}{5})$ during SrTiO$_3$ homoepitaxy at $T=970$ K, $P_{O_{2}}=10^{-5}$ Torr and $f=0.1$ Hz.  An absorber attenuated the specular beam.  Inset: Diffuse intensity during deposition of first half-monolayer.  Each curve represents the integrated signal following a pulse.
}
\label{ccd}
\end{figure}

To obtain information regarding morphological evolution, we measured the Small-Angle X-Ray Scattering (SAXS) during PLD (Fig.~\ref{ccd}).  A CCD detector recorded both the x-rays reflected specularly from the surface, and those with a small in-plane component, scattered in a narrow angular range about the specular beam.  Because the specular intensity dwarfed the diffuse by three orders of magnitude, we attenuated the specular with an aluminum absorber so that both signals could be simultaneously measured without saturating the CCD detector.

To increase the diffuse signal, we measured the x-rays scattered not around the often-used ``anti-Bragg'' $(00\frac{1}{2})$ position, but rather around the $(00\frac{1}{5})$ position of the specular Crystal Truncation Rod (CTR).  In general, the specular and diffuse signals intensify as one moves along the CTR toward $(000)$, although the intensity profile depends strongly upon the surface morphology \cite{alsnielsen}.  In layer-by-layer homoepitaxy, the specular intensity along the entire CTR oscillates once as each new layer is grown.

During layer-by-layer SrTiO$_3$ (STO) homoepitaxy, we observed that the diffuse and specular signals oscillate at the same frequency, out of phase (Fig.~\ref{ccd}). During growth of the first monolayer, the diffuse peaks are separated by $\Delta q_{\parallel}\!\approx\!0.06\,$\AA$^{-1}$ (Inset to Fig.~\ref{ccd}).  This indicates that surface features with a characteristic length scale of $\sim\!200\,$\AA~develop as the growing layer nears half-completion, and vanish at layer completion.  This length scale increases with increasing film thickness, eventually preventing resolution of the diffuse and specular signals. 

We interpret these data to indicate the nucleation of unit-cell high islands separated by $\sim\!50$ unit cells ($a_{STO}=3.905$~\AA).  This morphology is consistent with 20\,nm round features visible in our atomic force microscope (AFM) images and those in the literature~\cite{tischler}.  The nearly constant value of $q_{\parallel}$ (during the first monolayer) of the diffuse peaks indicates that while the islands grow in size as new material arrives, little subsequent nucleation occurs after the first pulse.  As islands coalesce, the diffuse signal corresponds to holes in the surface that fill in as the specular intensity approaches a maximum.

From the specular oscillations of Fig.~\ref{ccd}, we observe that growth proceeds at roughly one monolayer (ML) per $100\,$s.  The laser repetition rate ($f\!=\!0.1\,$Hz) and coverage of deposited material ($\sigma_{0}\!\approx\!0.1\,\mathrm{\frac{ML}{pulse}}$) set this time scale.  The shorter time scales associated with material incorporation and diffusion are regulated by the substrate temperature $T$, surface morphology, and diffusion energy barrier $U$ (and possibly also by the kinetic energy of incident particles ~\cite{fleet}).  While a rate-equation model~\cite{cohen,tischler} provides much qualitative insight regarding diffusion, the following continuum model allows direct extraction of $U$.

In a continuum model of diffusion, the surface density $\rho$ of diffusing species evolves according to the diffusion equation: $\frac{\partial \rho}{ \partial t}=D\nabla^{2} \rho\;$. The diffusion coefficient, $D$, depends on $T$ and $U$, via $D\!\propto\!e^{-U/k_{B}T}$~\cite{BCF}.  The Fourier transform of the diffusion equation yields $\rho_{q}\!\propto\!e^{-Dq^{2}t}$, where $q$ corresponds to the spatial frequencies present in $\rho$. In a boundary value problem, one expects a discrete spectrum of rate constants $k\!\propto\!-Dq^{2}$.  Short wavelength modes decay rapidly, and the spatial geometry determines the $k$-spectrum.  We emphasize that a single energy $U$ gives rise to multiple diffusion rates $k$.

The combined SAXS (Fig.~\ref{ccd}) and AFM data suggest a film morphology of round islands and holes.  We therefore solve the diffusion equation in a circular region of the (non-miscut) substrate of radius $r_{b}$, with a step located at $r_{a}<r_{b}$.  We write the solution as $\rho(r,\phi,t)=X(r,\phi)e^{-kt}$, and specify $X_{<}$ and $X_{>}$ as the spatial solutions in the regions $0\!<r\!<r_{a}$ and $r_{a}\!<r\!<r_{b}$, respectively.  Assuming $\frac{\partial \rho}{\partial \phi}=0$, $X$ takes the form of Bessel, $J_{0}(r)$, and Neumann, $N_{0}(r)$, functions.

We assume the step acts as a perfect sink, providing a boundary condition $\rho(r_{a})\!=\!0$~\cite{BCF,petrich}.  This method neglects the possible energy barrier for particles to cross downward over the step.  One can incorporate this ``Ehrlich-Schwoebel barrier'' by specifying the particle current $\vec{j}=-D\vec{\nabla}\rho$ at the boundary~\cite{tersoff,ghez}, without changing the conclusions of this paper.

The $X_{<}$ exclude the $N_{0}(r)$, which diverge as $r\!\rightarrow\!0$:
\[X_{<}=\sum_{m}A_{m}J_{0}(\alpha_{m}r)\equiv\sum_{m}X_{m<}\; .\]
Here $\alpha_{m}r_{a}$ is the $m$th root of $J_{0}(r)$.

We further assume that $\frac{\partial \rho}{\partial r}\arrowvert_{r_{b}}=0$, i.e. that the system boundary is far from any steps.  Then \[X_{>}=\sum_{m}B_{m}J_{0}(\beta_{m}r)+C_{m}N_{0}(\beta_{m}r)\equiv\sum_{m}X_{m>}\;,\]
where the boundary conditions determine the $\beta_{m}$ and the ratio $\frac{B_{m}}{C_{m}}$.  Finally, we specify $\rho(t\!=\!0,r)=\sigma_{0}$, where $\sigma_{0}$ is the coverage of single pulse.  This sets $A_{m}$ and $B_{m}$ via the orthogonality of the $J_{0}(r)$ and $N_{0}(r)$.

Kinematic theory accurately describes the scattered intensity.  For the simple case of momentum transfer normal to the surface of a non-miscut crystal, one has, in homoepitaxy \cite{ocko_91},

\begin{equation}
I(t)\propto\left | F \left ( q \right ) \right |^2 \,\left | \,\frac { 1 } { 1 - e^{i q d }}+\sum_n \, \theta_n \left ( t \right ) \, e^{- i q n d }\, \right |^2\; .
\label{eq:scatt_amp}
\end{equation}
Here $q$ is the scattering vector, $F \left ( q \right )$ is the scattering amplitude of a single layer, $\theta_n(t)$ is the time-dependent coverage of the $n^{th}$ layer, and $d$ is the layer spacing.  The first term represents the scattering from an ideally terminated single crystal.  The second term represents the scattering from the deposited film.  At the anti-Bragg position, x-rays scattered from adjacent layers interfere destructively, providing maximum sensitivity to single step height fluctuations.  Researchers often exploit this property in reflection high-energy electron diffraction (RHEED)\cite{karl,lippmaa_apl,blank,ohtomo_02} and x-ray growth studies~\cite{tischler,lee,fleet}.

Combining the complete solution $\rho(r,t)$ with Eq.~(\ref{eq:scatt_amp}), we calculate the anti-Bragg x-ray intensity, $I(t)$:
\begin{equation}
I(t)\propto ( \frac{1}{2} - \theta_{0} - \sigma_{0}+ 2\sum_{m}\sigma_{m}e^{-Dq_{m}^{2}t}\,)\,^{2},
\label{eq:int_diff}
\end{equation}
where $\sigma_0$ is the coverage of single pulse.  The pre-pulse surface coverage, $\theta_{0}$, the coverages $\sigma_{m}$ of species in the mass-losing layer, and the choice $q_{m}=\alpha_{m}$ or $\beta_{m}$ depend on whether the step at $r_{a}$ bounds an island or a hole.  The complicated time dependence of Eq.~(\ref{eq:int_diff}) involves multiple rate constants $k_{m}=Dq_{m}^{2}$, rather than a single exponential.

\begin{figure}
\resizebox{85mm}{!}{\includegraphics{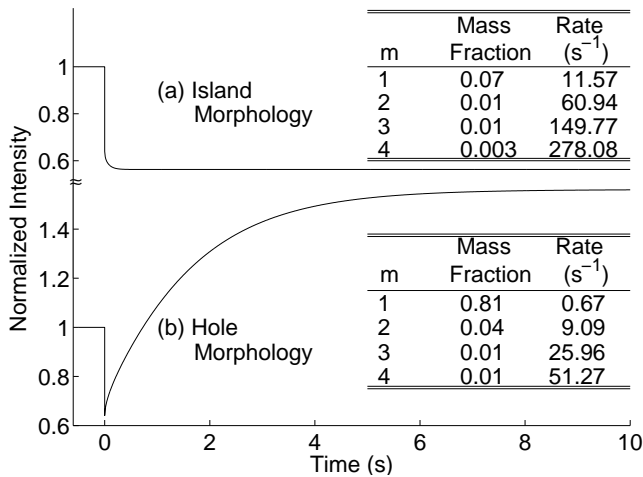}}

\caption{
Calculated anti-Bragg x-ray intensity following a laser pulse, using Eq.~(\ref{eq:int_diff}) and parameters in agreement with experiment.  The normalized intensity drops instantaneously upon pulse arrival, and then either falls (a) or rises (b), depending upon surface morphology.  Tables in each case list fractions of total deposited mass, $\frac{\sigma_{m}}{\sigma_{0}}$, and diffusion rates of the first four modes.  Integral forms of $\sigma_{m}$ were evaluated numerically.
}
\label{diffcurves}
\end{figure}

At the nucleation of each new layer, the morphology is characterized by islands.  In this regime, $\theta_{0}=(\frac{r_{a}}{r_{b}})^2$, $\sigma_{m}=\frac{2}{r_{b}^{2}}\int_{0}^{r_{a}}X_{m<}(r)rdr$, and $q_{m}=\alpha_{m}$.  Only species transferring off of the island affect $I$, so that $X_{>}$ does not influence the time-dependence.

In this case, $I$ slightly decreases following the large instantaneous drop due to pulse arrival [Fig.~\ref{diffcurves}(a)].  The growing layer moves toward half-completion and a minimum in the anti-Bragg oscillation.  The diffusion-moderated change is small, since most of the deposited material does not land on the island.  The small fractions of total deposited material, $\frac{\sigma_{m}}{\sigma_{0}}$, transferring between layers, move at relatively rapid rates $k_{m}$ [Inset to Fig.~\ref{diffcurves}(a)].  Many researchers have observed intensity transients in qualitative agreement with this form~\cite{tischler,blank}.

As the islands grow and coalesce, a network of holes better describes the surface.  Here, $\theta_{0}=1-(\frac{r_{a}}{r_{b}})^2$, $\sigma_{m}=\frac{2}{r_{b}^{2}}\int_{r_{a}}^{r_{b}}X_{m>}(r)rdr$, and $q_{m}=\beta_{m}$.  The intensity rises substantially following the instantaneous drop, as the film moves closer to the layer completion condition and growth oscillation maximum [Fig.~\ref{diffcurves}(b)].  The $k_{m}$ are in general slower than in the island case.  Roughly 80\% of deposited material is in the mode associated with $X_{0>}$, with a time constant $\frac{1}{k_{0}}$ equal to a few seconds [Inset to Fig.~\ref{diffcurves}(a)]. Of the species transferring into the hole, $\sim10\%$ move at rates $k_{m}\gg k_{0}$, adding faster components to the intensity change, noticeable at early times.

\begin{figure}
\resizebox{85mm}{!}{\includegraphics{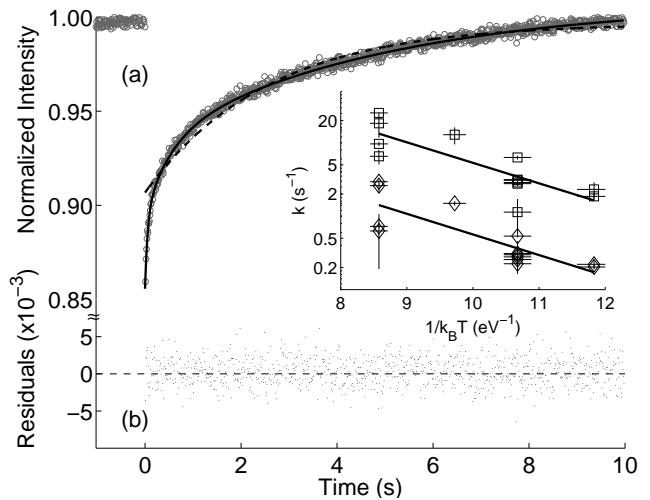}}
\caption{(a) Fits of Eq.~(\ref{eq:diffit}) to 900 K data with one (dashed curve) and three (solid curve) rate constants.  The circle size roughly equals the error of each point.  (b) Residuals of the fit with three rate constants.  Inset:  Arrhenius plot of 13 samples, with $T=900\rightarrow1060\,$K, and $P_{O_{2}}=10^{-6}\,$Torr.  Ordinate is $k_{1}$ (diamonds) and $k_{2}$ (squares) of Eq.~(\ref{eq:diffit}).}
\label{diffit}
\end{figure}

To test the model, we measured the anti-Bragg intensity during STO homoepitaxy via PLD (Fig.~\ref{diffit}), under conditions which promote steady-state layer-by-layer growth~\cite{song}.  Our PLD chamber is an integral component of a fully featured x-ray diffractometer that is permanently installed in the G3 hutch of the Cornell High Energy Synchrotron Source (CHESS)~\cite{fleet}.  By controlling the chamber oxygen pressure, $P_{O_{2}}$, substrate temperature, $T$, and laser repetition rate, $f$, we can select different growth modes, while monitoring film growth \emph{in-situ} with x-rays.  We grew STO films on STO (001) substrates with $P_{O_{2}}\!=\!10^{-6}\,$Torr, $T\!=\!900\!\rightarrow\!1060\,$K, and $f\!=\!0.1\,$Hz.  We set the laser energy density at the single crystal STO target to $2\!-\!3\,\mathrm{\frac{J}{cm^{2}}}$, and the target-substrate distance to 6\,cm, for an average $\sigma_{0}\!\approx\!0.1\,\mathrm{\frac{ML}{pulse}}$.  To prepare a TiO$_{2}$-terminated surface, we etched the substrates in buffered NH$_{4}$F-HF  \cite{kawasaki,lippmaa_ass}.  We annealed each sample in 1 mTorr O$_{2}$ for 1-2 hours at 1060 K prior to growth to obtain a smooth surface \cite{lippmaa_ass}.

To obtain the required time resolution, we recorded $I(t)$ with a multi-channel scaler (MCS) set to 10 ms dwell time per channel.  The MCS triggered the laser at the midpoint of each pass, collecting 5 s of intensity before and after each laser pulse.  We then normalized each MCS pass by $I(t)$ integrated over the pre-pulse region~\cite{fleet}.

We typically measured anti-Bragg intensities of ${10^4\mathrm{\frac{cts}{s}}}$.  This signal strength yields $10\%$ statistical noise in each MCS bin, comparable to the signal change $\frac{\Delta I}{I}$.  To reduce the noise, we averaged normalized MCS passes for a given sample over the entire deposition~\cite{fleet}.  While improving counting statistics, this method sacrifices information by combining intensity transients from a range of surface morphologies.  For example, during the growth of the first half-monolayer, individual transients resemble the curve of Fig.~\ref{diffcurves}(a)~(noise prevents quantitative analysis).

However, sums of MCS passes at constant growth oscillation phase reveal the transient to resemble the curve of Fig.~\ref{diffcurves}(b) at all phases~\cite{fleet}. This suggests that a hole morphology dominates throughout much of the growth.  A numerical simulation of a system containing an island and a hole reveals that the holes dictate the time structure of the intensity transient.  One can crudely predict this result by summing the two curves of Fig.~\ref{diffcurves}. Numerous authors have observed rising intensity transients during the ``falling'' portion of the anti-Bragg oscillation, where one might expect an island morphology ~\cite{karl,lippmaa_apl,blank}.

To fit the data, we truncated the series of Eq.~(\ref{eq:int_diff}) at $m=3$, for which the model predicts the $k_{m>3}$ to be much greater than $f_{\mathrm{Nyquist}}=50$ Hz.  In the hole regime, we have $\sigma_{m}\ll1$.  Retaining small quantities to first order,
\begin{equation}
I(t) = a - \sum_{m=1}^{3}b_{m}e^{-k_{m}t}\,.
\label{eq:diffit}
\end{equation}

We fit Eq.~(\ref{eq:diffit}) to data from 13 samples grown under layer-by-layer conditions.  For samples grown at $T\!>\!970\,$K, we required only two rate constants, suggesting that $k_{3}\gg f_{\mathrm{Nyquist}}$ at high T.  A typical fit [Fig.~\ref{diffit}(a)] describes the data well, with $\chi_{\nu}^{2}=1.1$ and normally distributed residuals [Fig.~\ref{diffit}(b)].  Adding terms $m\!>\!3$ did not improve $\chi_{\nu}^{2}$.   For comparison, we also show a fit with the series truncated at $m=1$, a form often assumed for the intensity evolution~\cite{karl,blank}.  This simple exponential clearly fails to describe the data at early times ($\chi_{\nu}^{2}=4.9$).

Averaged over all samples, the ratio $\frac{k_{2}}{k_{1}}\!=\!11.9$ agrees with the hole model prediction of 9 to 13, which can vary depending on hole size and shape.  The ratio $\frac{k_{3}}{k_{1}}=80$ exceeds the hole model prediction of 30.   This may be due to a small island contribution, for which $\frac{k_{2,island}}{k_{1,hole}}\approx100$.  Additionally, the fixed $r_{a}$ approximation used to derive Eq.~(\ref{eq:int_diff}) decreases the ratio $\frac{k_{m>1}}{k_{1}}$.  In reality, the holes fill in, effecting longer saturation times for the x-ray intensity, and reducing the measured $k_{1}$.

The ratio $\frac{b_{1}}{b_{2}}\approx2$ is smaller than the model prediction of 10. High frequency terms thus contribute more to the signal than predicted.  This discrepancy could arise from the approximation of a two-level system.  The actual broad growth interface would place more material near step-edges, increasing the rate of the intensity change.

In the inset to Fig.~\ref{diffit}, we present $k_{1}$ and $k_{2}$ in an Arrhenius plot for the 13 samples mentioned above.  Since $k_{m}\propto D$, the slopes of linear fits give the diffusion activation energy, $U$.  The slopes of separate fits to the two distributions agree to within 1\%, revealing that both rates arise from a single energy.  A simultaneous fit of both distributions yields $U=0.6\pm0.2$ eV, in the range of published diffusion barriers for oxides~\cite{harris}.  One source of error, due to sums over multiple MCS passes could be eliminated by higher x-ray intensity, allowing analysis of individual laser pulses.

That a single process generates multiple time scales in the diffracted intensity is a direct consequence of the characteristic length scale on the surface, revealed in our SAXS data.  The presence of this length scale forces a discrete $q$-spectrum upon the density of diffusing species.  Higher-$q$ components decay quickly as particles diffuse to steps, effecting a rapid early response in the time-resolved anti-Bragg intensity.  This effect is more easily seen in PLD than in steady state growth techniques, due to the high density immediately following a laser pulse.

This work is supported by the Cornell Center for Materials Research, under National Science Foundation (NSF) Grant No.~DMR-0079992.  This research used the G-line facilities at CHESS.  The construction of the G-line facility was supported by the NSF under Grant No.~DMR-9970838.  CHESS is supported by the NSF and the NIH/NIGMS under Grant No.~DMR-0225180.
\bibliography{sources}
\end{document}